\documentclass[a4paper, 11pt]{article}
\RequirePackage{amsmath}
\RequirePackage{amssymb}
\RequirePackage{epsfig}
\RequirePackage{graphicx}
\RequirePackage{color}
\RequirePackage[colorlinks=true
,urlcolor=blue
,anchorcolor=blue
,citecolor=blue
,filecolor=blue
,linkcolor=blue
,menucolor=blue
,linktocpage=true
,pdfproducer=medialab
,pdfa=true
]{hyperref}
\usepackage{authblk}
\usepackage[margin=1 in]{geometry}

\begin{document}
\title{On DUNE prospects in the search for sterile neutrinos} 
\author[1]{Igor Krasnov}
\affil[1]{Institute for Nuclear Research of the Russian Academy of Sciences, 60th October Anniversary Prospect 7a, 117312, Moscow, Russia}
\affil[1]{\href{mailto:iv.krasnov@physics.msu.ru}{iv.krasnov@physics.msu.ru}}

\maketitle

\abstract{
Experiments measuring the parameters of active neutrino oscillations can also search for the sterile neutrinos in a part of sterile neutrino parameter space.
In this paper, we analyze the prospects for the sterile neutrino search in the upcoming experiment DUNE for the sterile neutrinos with masses at GeV scale.
As it relies on the still-undecided design of the Near Detector, we provide the expected number of the sterile neutrino decays in the Near Detector volume.
Our most optimistic predictions show that the corresponding limit on mixing can be approximately of the same order as the previous estimates made for the LBNE.
We present our results as separate plots for the sterile neutrino mixing with electron, muon and tau neutrinos.
Generally, DUNE has good prospects to probe large region of the previously unavailable part of the parameter space before the new projects (like SHiP) join the searches.
}

\section{Introduction}

Physics beyond the Standard Model (SM) of particle physics is one of the most rapidly developing fields in theoretical physics.
It stems from the discrepancies between SM predictions and some of the experimental data, obtained in the last decades.
For example, neutrino oscillation phenomena show that the SM is not complete.
One way to address this problem is to introduce additional leptons, sterile with respect to the SM gauge interactions $SU(3)_c \times SU(2)_W \times U(1)_Y$ \cite{Mohapatra:1979ia}.
They are usually called sterile neutrinos and introduced in the following way:
\begin{equation}
\label{eq1}
\mathcal{L}=\mathnormal{i} \bar{N}_I\gamma^\mu \partial_\mu N_I-\Big(\frac{1}{2} M_I \bar{N}^c_I N_I + Y_{\alpha I}\bar{L}_\alpha \tilde{H} N_I +h.c.\Big),
\end{equation}
here $N_I$ are sterile neutrinos, $M_I$ are their Majorana masses, and $Y_{\alpha I}$ stand for their Yukawa couplings with lepton doublets $L_\alpha, \alpha= e, \mu, \tau$ and SM Higgs doublet ($\tilde{H}_a=\epsilon_{a b} H^*_b$).
One needs at least two sterile neutrinos to explain active neutrino oscillations, and at least three in the case when all active neutrinos have non-zero masses.
It was shown that heavy sterile neutrinos may also provide an explanation for leptogenesis (see, for example, Ref. \cite{Boyarsky:2009ix}) or serve as a dark matter candidate \cite{Adhikari:2016bei}.

The strategy of the search for such particles depends heavily on their masses.
If sterile neutrinos have masses at GeV scale, they can appear in heavy hadron decays.
Such sterile neutrinos can be searched for in various collider experiments.
Experiments measuring the parameters of neutrino oscillations are capable of detecting sterile neutrino decay events as well.
Beam energies, as well as specifics of measurement processes, geometry and the relative position of the detector, determine the region of sterile neutrino parameter space that can be tested in a given experiment.
Experiments such as CHARM \cite{Bergsma:1985is}, NuTeV \cite{Vaitaitis:1999wq}, PS191\cite{Bernardi:1987ek}, DELPHI \cite{Abreu:1996pa}, OKA \cite{Sadovsky:2017qsr}, LHCb \cite{Antusch:2017hhu,Canetti:2014dka}, Belle \cite{Canetti:2014dka}, E949 \cite{Artamonov:2014urb} provide limits on active-sterile neutrino mixing.
Many developing projects and upcoming experiments, such as NA62 \cite{CortinaGil:2017mqf,Drewes:2018irr}, SHiP \cite{SHiP:2018xqw}, MATHUSLA \cite{Curtin:2018mvb},  T2K \cite{Izmaylov:2017lkv} and DUNE \cite{Acciarri:2016crz,Acciarri:2015uup,Strait:2016mof,Acciarri:2016ooe} declare the search for heavy neutral leptons to be one of their goals.

The search for the sterile neutrinos in beam-dump experiments was, for example, considered in Refs. \cite{Gorbunov:2007ak, Gninenko:2013tk}.
The process behind the search can be described as follows: the proton beam strikes the target and produces a great number of heavy secondary mesons.
Due to active-sterile neutrino mixing a part of these mesons would produce sterile neutrinos in their decays.
A part of these sterile neutrinos flies towards the detector and decays inside its volume.
Such decays can be observed.

The LBNE project provided their estimate of active-sterile neutrino mixing by rescaling the results of existing experimental data using the new experiment specifics in their design report \cite{Adams:2013qkq}.
The DUNE project inherited this estimate as their own predicted sensitivity to active-sterile neutrino mixing without updating the specifics of the experiment such as the Near Detector length.
Until now there was no update made for the proposed DUNE Near Detector design \cite{Acciarri:2016crz,Acciarri:2015uup,Strait:2016mof,Acciarri:2016ooe}.

The aim of this paper is to calculate the sensitivity of DUNE to the active-sterile neutrino mixing for sterile neutrinos of masses at GeV scale.
As it relies on yet to be decided design of the Near Detector, it is impossible to provide a proper estimation of the number of background events.
In this paper we present the iso-contours for the number of expected heavy neutrino decays inside the detector volume, in the plane $M_N - |U|^2$, thus avoiding the issue of dealing with the experimental detection efficiencies and reconstruction effects.
We also propose some ideas as to how it might be possible to enhance the signal to background ratio.
This paper can be useful for the consideration of the DUNE Near Detector design or its possible additional upgrades.

The paper is organized as follows.
In Sec. \ref{sec:DUNE} we present the overall layout of DUNE near detector facilities and the relevant proton beam properties.
After that in Sec. \ref{sec:features} we list the experimental features of the search, such as different meson production rates and their momentum distribution.
We present in more detail the analysis of sterile neutrino detection specifics and our way to account for it in Sec. \ref{sec:algorithm}.
We present our estimates in Sec. \ref{sec:results}, and some possible issues in Sec. \ref{sec:background}.
We conclude in Sec. \ref{sec:conclusions}.
We also present the various relevant experimental data in Appendix \ref{sec:parameters} and the sterile neutrino-related formulae in Appendix \ref{sec:formulae}.

\section{DUNE}
\label{sec:DUNE}
The main goal of DUNE is to measure active neutrino parameters with high precision \cite{Acciarri:2016crz,Acciarri:2015uup,Strait:2016mof,Acciarri:2016ooe}.
This will be achieved by creating very intensive high-energy neutrino flux.
High energy proton beam (up to 120 GeV) strikes a target, producing a high number of secondary particles (mainly pions and kaons) which can produce sterile neutrinos during their decay.
To provide enough space for the secondary particles to decay, a 221~m long and 4~m wide decay pipe is planned to be installed behind the target area.
At the end of the pipe, the absorber is placed to reduce the background from muons.
Additionally, natural rock fills the area between the decay pipe and the detector.
The resulting neutrino beam is directed towards the Near Detector at 574 m from the target and the Far Detector at 1300 km, which allows for better prospects for active neutrino parameters measurement.

Important properties of a reference proton beam are listed in Tab. \ref{tab:proton}.
Geometrical sizes are listed in Tab. \ref{tab:geometry}.
\begin{table}[h]
\begin{center}
\begin{tabular}{| p{6 cm} | p{2 cm} |}
\hline
Proton beam energy & 120 GeV\\
\hline
Spill duration & $1.0 \times 10^{-5}$ s\\
\hline
Protons on target per year & $1.1 \times 10^{21}$\\
\hline
Cycle time & 1.2 s\\
\hline
\end{tabular}
\caption{Proton beam properties \cite{Strait:2016mof}.}
\label{tab:proton}
\end{center}
\end{table}

\begin{table}[h]
\begin{center}
\begin{tabular}{| p{8.5 cm} | p{4 cm} |}
\hline
Distance from the target to the Near Detector $L$ & 574 m\\
\hline
Decay pipe length $l_{decay\,pipe}$ & 194 m\\
\hline
Decay pipe radius $r_{decay\,pipe}$ & 2 m\\
\hline
Near Detector reference size $\Delta l \times \Delta h \times \Delta h$ \cite{Acciarri:2016ooe}& $6.4$ m $\times 3.5$ m $\times 3.5$ m\\
\hline
\end{tabular}
\caption{Geometrical sizes \cite{Strait:2016mof}.}
\label{tab:geometry}
\end{center}
\end{table}

As the Near Detector is located considerably far from the target, one can notice that the heavy sterile neutrino would reach the detector later than the active neutrinos produced at the same time.
But that shift in arrival time is generally less than active neutrino travel time from target to the detector $t_\nu \approx  \frac{ 574 m}{3 \times 10^8 m/s} = 1.91$ $\mu$s.
And the latter time is considerably less than the spill duration of $\tau = 10 \mu$s.
We can use a timing cut only if the sterile neutrino arrives at the detector after the last of the active neutrinos, produced during the spill, passed through it.
There are two random elements in this.
First of all, if we start time count at the beginning of the spill, the moment of the possible sterile neutrino production $t_0$ is evenly distributed between 0 and $\tau$.
The other random element is the value of the sterile neutrino momentum.
We present the way we construct the sterile neutrino momentum distribution in Sec. \ref{sec:algorithm}.
The timing cut criterion for the sterile neutrino with velocity $v_N$ can be expressed as follows: $ t_0 + \frac{L}{v_N} > \tau + \frac{L}{c}$.
Therefore the application of the timing cut is more probable for the sterile neutrinos produced at the end of the spill, but it also can be applied to the sterile neutrinos that were produced early if their momentum is sufficiently small.
Timing cut can be used for all sterile neutrinos with mass $M_N$ and momentum $p<\frac{L}{\sqrt{(L+c\tau)^2-L^2}} M_N$.
The probability $P$ for the timing cut to apply to the produced sterile neutrino with mass $M_N$ and fixed momentum $p>\frac{L}{\sqrt{(L+c\tau)^2-L^2}} M_N$ reads: $P = \frac{L}{c \tau} (\sqrt{1+\frac{M_N^2}{p^2}}-1)$.
Taking into account the sterile neutrino momentum distribution (see Sec. \ref{sec:algorithm}), we obtain that less than $0.1\%$ of the sterile neutrinos that fly in the direction of the detector satisfy the timing cut criterion.
This means that, for the most part, the active neutrino spill would overlap with the sterile neutrino arrival, serving as a background for the sterile neutrino search.
We note that the active neutrinos are generally produced in the decays of pions or kaons.
These mesons have considerable mass, and that means their travel time before their decay, compared to our estimate, contains an additional delay in the active neutrino arrival time.
The same reasoning applies to the sterile neutrinos produced in the kaon decays, which results in the additional delay in the sterile neutrino arrival time for these sterile neutrinos, although that delay is shorter than the delay for the active neutrinos produced in pion decays.
Therefore our estimate is conservative.
Summing up, timing doesn't help to get rid of the background from active neutrino interactions in the Near Detector.
However, a special run with much shorter spill duration and lower beam energy can be considered as a solution to this problem.

The Far Detector is simply too far to provide a sufficient number of sterile neutrino decays in the detector volume.
We discuss geometrical restrictions in Sec. \ref{sec:algorithm}.

\section{Experimental features}
\label{sec:features}
We assume that the primary 120 GeV proton beam strikes a target and produces mesons that may decay into the sterile neutrinos and SM particles.
Sterile neutrino momentum $p_N$ and energy $E_N$ spectra are very important for further analysis.
They are closely related to the momentum $p_H$ and energy $E_H$ spectra of secondary mesons.
Each momentum has the longitudinal $p_L$ and transversal $p_T$ components.
The longitudinal component is the one directed along the axis of the active neutrino beam, and the transversal component is orthogonal to the longitudinal one.

It is shown in Ref. \cite{Gorbunov:2007ak}, that the number $dN_H$ of heavy hadrons is proportional to the differential cross section of relevant hadrons direct production $d\sigma_H$:
\begin{equation}
\label{meson_momentum}
\frac{dN_H}{dp_{H_L}dp^2_{H_T}} \propto \frac{d\sigma_H}{dp_{H_L}dp^2_{H_T}}.
\end{equation}

The distribution of the longitudinal momentum of secondary mesons can be fitted from experimental data, with fit parameters varying with beam energy.
The usual approximation for the longitudinal momentum $p_{H_L}$ distribution of the differential cross section $d\sigma_H$ reads \cite{Kodama:1991jk,AguilarBenitez:1987rc}:
\begin{equation}
\label{meson_momentum_L}
\frac{d \sigma_H}{d x_F} \propto \left( 1 - x_F \right)^c, x_F = \frac{p_{H_L}}{p_{H_L}^{max}},
\end{equation}
with $c = 3$ being the phenomenological factor-of-two estimate for the relevant energy $E = 120$ GeV (see \cite{Gorbunov:2007ak}).

The distribution of the transversal momentum of secondary mesons depends heavily on  the details of hadronization \cite{Gorbunov:2007ak}.
They are usually approximated with fragmentation function $D(z)$.
We take PYTHIA distributions i.e. the Lund fragmentation function \cite{PYTHIA}:
\begin{equation}
\label{Lund}
D(z)= \frac{(1-z)^a}{z^{1+ r_Q b m_Q^2}}\exp\left(- \frac{b}{z} (M_H^2+p^2_{H_T})\right),
\end{equation}
where $z$ represents a part of hadron momentum $p_H$ carried by heavy quark $p_Q$.
The default parameter values (the ones we take) are $a=0.68, b=0.98$ GeV$^{-2}, r_s=0, r_c=1.32, r_b=0.855$ \cite{PYTHIA}.
Heavy quark masses are $m_c = 1.275$ GeV, $m_b = 4.18$ GeV \cite{Tanabashi:2018oca}.
The resulting transversal momentum distribution of secondary mesons reads:
\begin{equation}
\label{meson_momentum_T}
\frac{d\sigma_H}{dp^2_{H_T}} \propto \int_0^1 dz \frac{(1-z)^a}{z^{1+ r_Q b m_Q^2}}\exp\left(- \frac{b}{z} (M_H^2+p^2_{H_T})\right).
\end{equation}

Different mesons have different chances to be produced in a specific experiment.
This is determined by two factors: how many quarks of the corresponding type $\chi_q$ is generated by interactions of the primary beam with the target and the weight of a specific channel in quark hadronization $Br(q \to H...)$.
Basically after the number of simpifications Eq. \eqref{meson_momentum} is transformed into the following form \cite{Gorbunov:2007ak}:
\begin{equation}
\label{eq:N_H}
N_H = N_{POT} \times M_{pp} \times \chi_q \times Br(q \to H),
\end{equation}
where $N_H$ is a number of secondary hadrons and $N_{POT}$ is a total number of ``protons on target'' (we identify it with the total number of proton interactions in the thin target).
$M_{pp}$ is the multiplicity of reaction,  i.e. the number of the hadrons produced in the interaction of primary protons with the target.
It is already taken into account for all considered mesons except K-mesons in the value of $\chi_q$.
We take $M_{pp}=1$ for these mesons, for K-mesons the value $M_{pp} > 1$ depends on the primary beam energy: $M_{pp}(K)=11$ for $E=120$ GeV \cite{Gorbunov:2007ak}.
We take the following values of $\chi_q$ \cite{Lourenco:2006vw}:
\begin{equation}
\label{chi}
\chi_s \equiv \frac{\sigma_{pp \to s}}{\sigma_{pp_{total}}} = \frac{1}{7}, \; \chi_c \equiv \frac{\sigma_{pp \to c}}{\sigma_{pp_{total}}} = 10^{-4}, \; \chi_b \equiv \frac{\sigma_{pp \to b}}{\sigma_{pp_{total}}} = 10^{-10}.
\end{equation}
For $s$-quark production fractions we take \cite{Gorbunov:2007ak}:
\begin{equation}
\label{Br_s}
Br (s \to K^-) = Br (s \to K^0_L) = Br (s \to K^0_S) = 1/3.
\end{equation}
For $c$-quark production fractions we take \cite{SHiP:2018xqw}:
\begin{equation}
\label{Br_c}
Br(c \to D^+) = 0.207,\; Br (c \to D^0) = 0.632,\; Br (c \to D^+_s) = 0.088.
\end{equation}
For $b$-quark production fractions we take \cite{Tanabashi:2018oca}:
\begin{equation}
\label{Br_b1}
Br (b \to B^+) = Br (b \to B^0) = 0.405, \; Br (b \to B_s^0) = 0.101.
\end{equation}
The production fraction $Br (b \to B_c^+)$ has only been measured at LHC energies, where it reaches few $\times 10^{-3}$ \cite{Bondarenko:2018ptm}.
At lower energies, it is not known.
We take:
\begin{equation}
\label{Br_b2}
Br (b \to B_c^+) = 10^{-3}.
\end{equation}

Sterile neutrino production fractions from various mesons as well as sterile neutrino decay modes are listed in Appendix \ref{sec:formulae}.

\section{Algorithm}
\label{sec:algorithm}

\begin{figure}
\begin{center}
\includegraphics[width=0.95\linewidth]{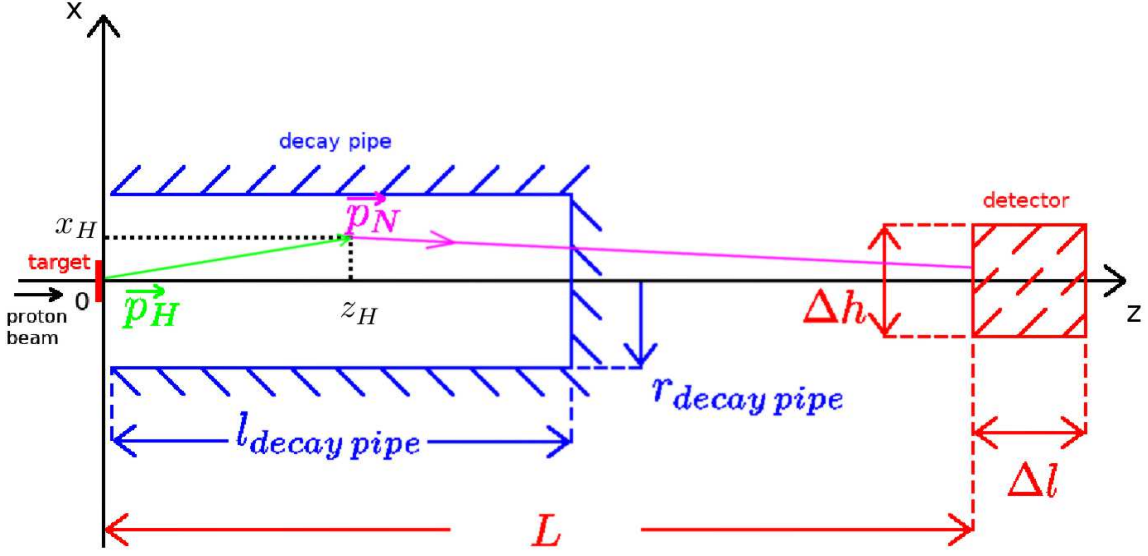}
\caption{Schematic illustration of the detector geometry.}
\label{fig:scheme}
\end{center}
\end{figure}

We want the detector to detect the sterile neutrino decay signals and distinguish them from the background.
The main question is: how small the value of $|U^2|$ can be to still allow it? 
The answer to the question depends heavily on the detector configuration, its efficiency and the methods used to reduce the background.
As none of the above can be clarified yet, we only provide the iso-contours for the number of expected heavy neutrino decays $N_{detector}$ inside the detector volume, in the plane $M_N - |U|^2$ (we plan to provide the proper estimation of the number of background events in another paper after the Near Detector design will be fixed).
We simply scan the values of $M_N$ with a 20 MeV step, starting from $M_N = 140$ MeV.
As we increase the value of $M_N$ we eventually reach the value for which the predicted number of expected heavy neutrino decays becomes less than the specified value $N_{detector}$ for a specific iso-contour.
We abort our scan at this mass value.

First, we calculate the energy distribution function in each process according to Eqs. \eqref{Br2}, \eqref{Br3} in Appendix \ref{app:production}.
Note that this is an energy in the rest frame of the decaying meson $H$. 
We also calculate the sterile neutrino mean lifetime $\tau_N = \frac{1}{\sum \Gamma (N \to ...)}$ according to Eqs. \eqref{eq:decay_first} -- \eqref{eq:decay_last} in Appendix \ref{app:decay}.

After that we randomly chose one of the processes, using corresponding weight $\chi_q \times Br(q \to H) \times Br (H \to N...)$ according to formulae \eqref{Br2}, \eqref{Br3}, \eqref{Br_vec}, \eqref{eq:tau1} -- \eqref{eq:tau3} from Appendix \ref{app:production}, and values \eqref{chi} -- \eqref{Br_b2}.
The following decays are significant: $K^0_L \to \pi^+ l^- N,\; K^+ \to \pi^0 l^+ N,\; D^0 \to K^+ l^- N,\; D^+ \to K^0 l^+ N,\; D^0 \to \pi^+ l^- N,\; D^+ \to \pi^0 l^+ N,\; D_s^+ \to \eta^0 l^+ N,\; B^+ \to D^0 l^+ N,\; B^0 \to D^+ l^- N,\; D^0 \to K^{+*} l^- N,\; D^+ \to K^{0*} l^+ N,\; B^0 \to D^{+*} l^- N,\; B^+ \to D^{0*} l^+ N,\; B_s^0 \to D_s^+ l^- N,\; B_s^0 \to D_s^{*+} l^- N$.
The indirect production of the mesons from the decays of heavier mesons can be neglected, as their number is several orders of magnitude smaller than the number of the mesons produced in proton collisions in the target, as can be seen from Eqs. \eqref{eq:N_H}, \eqref{chi}.
We randomly choose $E_N$ from the previously calculated distribution that corresponds to the chosen process.
From the distributions described by \eqref{meson_momentum}, \eqref{meson_momentum_L} we randomly choose $p_{H_L}, p_{H_T}, p_H^2 = p_{H_L}^2 + p_{H_T}^2$.

There is no preference for the direction of sterile neutrino momentum $p=\{p_x, p_y, p_z\}$ in the rest frame of $H$, so this direction is also chosen randomly.
Its absolute value is $p = \sqrt{E_N^2-M_N^2}$.
Then we make a Lorentz boost to this momentum to calculate the resulting sterile neutrino momentum $p_N$ in the laboratory frame.
We choose the target as the point of origin, $z$ axis is directed towards the detector and $x,y$ axes are chosen so that $p_{H_L} \equiv p_{H_z}, p_{H_T} \equiv p_{H_x}, p_{H_y}=0$.
For convenience, we plot the scheme of the process geometry in Fig. \ref{fig:scheme}.
The resulting longitudinal and transversal components of sterile neutrino momentum in the laboratory frame read:
\begin{eqnarray}
\label{p_Ny}
p_{N_x} & = & -\frac{E_N}{M_H} p_{H_T} - p_z \sqrt{1+\frac{p_H^2}{M_H^2}}\frac{p_{H_T}}{p_H} + p_x \frac{p_{H_L}}{p_H},\\
\label{p_Nz}
p_{N_y} & = & p_y,\\
\label{p_Nx}
p_{N_z} & = & \frac{E_N}{M_H} p_{H_L} + p_z \sqrt{1+\frac{p_H^2}{M_H^2}} \frac{p_{H_L}}{p_H} + p_x \frac{p_{H_T}}{p_H}\quad \equiv\quad p_{N_L}.
\end{eqnarray}

Note that according to \eqref{p_Nx} if decaying meson velocity in the laboratory frame $v_{H_{lab}} = \frac{p_H}{E_H}$ is smaller than sterile neutrino longitudinal velocity in the meson rest frame $v_{N_{L_H}} = \frac{p_z}{E_N}$, $v_{H_{lab}} < v_{N_{L_H}}$, then it is possible that $p_{N_z}<0$, i.e. sterile neutrino flies in the direction opposite of the detector.
Obviously, such sterile neutrino won't be detected.

After the proton beam strikes the target, produced secondary particles travel some distance away from the target before decaying.
The distance that $H$ meson travels down the pipe before decaying at the moment $t_H$ is $z_H=\frac{p_{H_L}}{M_H} t_H$ ($t=0$ corresponds to the moment when the proton beam strikes the target).
Its shift from the axis at this moment is $x_H = \frac{p_{H_T}}{M_H} t_H$.
For short living mesons $t_H \sim 0, z_H \sim 0, x_H \sim 0$.
If the meson produces sterile neutrino, its initial coordinates are $x_N(t_H)=x_H, y_N(t_H)=0, z_N(t_H)=z_H$.
One of the criteria for the sterile neutrino to decay in the detector volume is for it to decay when $L<z_N(t_H+\Delta t)<L+\Delta l$.
Here $L$ is the distance from the target to the detector, $\Delta l$ is the effective length of the detector and $\Delta t =\frac{M_N}{p_{N_L}} (L-z_H)$ is a time it takes for the sterile neutrino to travel the distance $L- z_H$.
As $\Delta l \ll L$, we simply can take $z_N(t_H+\Delta t)=L$.
The other sterile neutrino coordinates $x_N, y_N$ can be expressed as:
\begin{eqnarray}
\label{eq:yN}
x_N&=&\frac{p_{H_T}}{M_H} t_H + \frac{p_{N_x}}{p_{N_z}}\left( L - \frac{p_{H_L}}{M_H} t_H \right),\\
\label{eq:zN}
y_N&=&\frac{p_{N_y}}{p_{N_z}}\left( L - \frac{p_{H_L}}{M_H} t_H \right).
\end{eqnarray}
For the on-axis detector we take that the number of sterile neutrinos flying in the direction of the detector $\mathcal{N}_{forward}$ is the number of sterile neutrinos for which the following statement is true:
\begin{equation}
\label{criteria}
\sqrt{x_N^2+y_N^2} < \frac{\Delta h}{2}.
\end{equation}
Here $\Delta h$ is the transversal size (height and width) of the detector.
Equation \eqref{criteria} means that the sterile neutrino won't fly towards the detector if it deviates too much from the axis $z$.
For short living mesons $t_H \sim 0, z_H \sim 0$ Eq. \eqref{criteria} turns into:
\begin{equation}
\label{eq:geometry_simple}
\frac{p_{N_T}}{p_{N_L}} <\frac{\Delta h}{2 L},
\end{equation}
where $p_{N_T} = \sqrt{p_{N_x}^2+p_{N_y}^2}$ is the sterile neutrino transversal momentum.

Another random variable is the moment of decay of the meson $H$.
Probability for meson $H$ to decay before the moment $t_H$ in the meson rest frame is:
\begin{equation}
\label{eq:probability}
P\left(t_H\right) = 1 - exp\left(-\frac{t_H}{\tau_H}\right),
\end{equation}
where $\tau_H$ is the meson mean life-time.
We choose $t_H$ according to this law \eqref{eq:probability}.

For long-lived mesons (kaons) we have additional consideration: if the kaon longitudinal travel distance $z_H$ exceeds the decay pipe length $l_{decay\,pipe}$, $z_H > l_{decay\,pipe}$, then the kaon reaches the absorber.
In pretty much the same way if the kaon transversal travel distance $x_H=\frac{p_{H_T}}{M_H} t_H$ exceeds the decay pipe radius $r_{decay\,pipe}$, $x_H>r_{decay\,pipe}$, then it collides into the decay pipe walls.
When either of these happens, the kaon usually rapidly loses energy.
At the moment of its decay it practically stops.
From Eqs. \eqref{eq:yN} -- \eqref{eq:geometry_simple} it is obvious that sterile neutrinos from such kaons are not very relativistic and have a very small probability to reach the detector hundreds of meters away.
For that reason, we consider the contribution of such kaons negligible and remove them from our estimates at the moment they reach the absorber or the walls. 

We repeat this process many times (we take $\mathcal{N}_{total} = 10^7$ iterations) and take into account only those events, that satisfy criterion \eqref{criteria} and obtain the number of sterile neutrinos $\mathcal{N}_{forward}$ that fly in the direction of the detector.
The portion of sterile neutrinos $\zeta_N$ that flies towards the detector reads:
\begin{equation}
\zeta_N = \frac{4}{\pi}\frac{\mathcal{N}_{forward}}{\mathcal{N}_{total}},
\end{equation}
where coefficient $\frac{4}{\pi}$ represents the fact that the frontal surface of the detector is a square and not a circle, as implied in \eqref{criteria}.

We also obtain the resulting distribution $f_1(p_{N_L})$ of the longitudinal momentum $p_{N_L}$ of the sterile neutrinos that fly in the direction of the detector, $\int f_1(p_{N_L})dp_{N_L}=1$.
We note that for the sterile neutrino mixing with tau neutrino a ``three stages'' processes become important, when heavy meson decays producing tauons, and the sterile neutrino is produced in tauon decays.
We discuss this case in more detail in Appendix \ref{app:production}.

The last distribution we need is the distribution $f_2(z_H, p_{N_L})$ of the kaon longitudinal travel distance $z_H$, $\int f_2(z_H, p_{N_L})dz_H=1$, which has to be taken into consideration for the sterile neutrinos produced in the kaon decays.

In the experiment, the total number of produced sterile neutrinos $N_N$ depends on the number of mesons of each type $N_H$ (see Eq. \eqref{eq:N_H}) produced at the target and the probability for them to produce the sterile neutrino $Br (H \to N...)$ (see Appendix \ref{app:production}).
It can be written as:
\begin{equation}
\label{eq:N_N}
N_N = N_{POT} \times \sum_{q,H} M_{pp}(H) \chi_q Br(q \to H) Br (H \to N...).
\end{equation}

The probability of sterile neutrino decay during the time $t_N$ after its production in the sterile neutrino rest frame is described by the usual law \eqref{eq:probability}, where meson $H$ is replaced by sterile neutrino $N$:
\begin{equation}
\label{eq:probability2}
P\left(t_N\right) = 1 - exp\left(-\frac{t_N}{\tau_N}\right),
\end{equation}
here $\tau_N = \frac{1}{\sum \Gamma (N \to ...)}$ is the sterile neutrino lifetime (see Eqs. \eqref{eq:decay_first} -- \eqref{eq:decay_last} in Appendix \ref{app:decay}).
If one wants to use any additional cuts that consider only specific channels, one would have to take into account only these channels.
We list some remarks on this possibility in Sec. \ref{sec:results}.

In the laboratory frame the sterile neutrino with mass $M_N$ and longitudinal momentum $p_{N_L}$ travels the distance $l_N = \frac{p_{N_L}}{M_N} t_N$ along the beamline before decaying.
We note that for the kaons we also have to take into account the distance $z_H = \frac{p_{H_L}}{M_H} t_H$ they travel in the decay pipe before decaying. 
The probability the sterile neutrino decays into the $\it{visible}$ modes in the interval $L < z_H + l_N < L +\Delta l$ is:
\begin{equation}
\label{eq:probability3}
P\left( L < z_H + l_N  < L +\Delta l \right) = exp\left(-\frac{L - z_H}{\tau_N}\frac{M_N}{p_{N_L}}\right) \left( 1 - exp\left(-\frac{\Delta l}{\tau'_N}\frac{M_N}{p_{N_L}}\right)\right).
\end{equation}
Here $(\tau'_N)^{-1}$ is the sum of all detectable ($\it{visible}$) sterile neutrino decay modes \eqref{eq:decay_first} -- \eqref{eq:decay_last}, i.e. all modes besides the three neutrino decay channel \eqref{eq:decay_3nu}.
If one needs to take into account only some specific decay modes, for example, to apply some background cut, one has to consider only these modes in $(\tau'_N)^{-1}$.

To account for the distribution of the sterile neutrino momentum $p_{N_L}$ and the distribution of the kaons longitudinal travel distance $z_H$ the resulting value of the sterile neutrino probability to decay in the detector volume reads:
\begin{equation}
\label{eq:probability_all}
P = \int_0^{p_{N_L}^{max}} dp_{N_L} f_1(p_{N_L}) \int_0^{l_{decay~pipe}} dz_H f_2(z_H, p_{N_L}) P\left( L < z_H + l_N  < L +\Delta l \right).
\end{equation}

The number of sterile neutrinos $N_{detector}$, that decay with the probability $P$ inside the detector of the length $\Delta l$, can be expressed as:
\begin{equation}
\label{eq:N_detect}
N_{detector} = N_N \cdot \zeta_N \cdot P,
\end{equation}
where $N_N$ is the total number of produced sterile neutrinos and $\zeta_N$ is the portion that flies towards the detector.
For the sterile neutrino with fixed mass and mixing, Eq. \eqref{eq:N_N2} gives us the total number of the sterile neutrino decays inside the detector volume.
Sterile neutrino lifetime depends on the mixing as $\tau_N = T_N |U|^{-2}$, where numerical coefficient $T_N$ doesn't depend on $|U|^2$  (see Eqs. \eqref{eq:decay_first} -- \eqref{eq:decay_last} in Appendix \ref{app:decay}).
In the same way $\tau'_N = T'_N |U|^{-2}$, numerical coefficient $T'_N$ doesn't depend on $|U|^2$.
For the number of produced sterile neutrinos we have $N_N = \mathcal{N}_N |U|^2$, where numerical coefficient $\mathcal{N}_N$ doesn't depend on $|U|^2$ (see \eqref{eq:N_N} and Eqs. \eqref{Br2}, \eqref{Br3}, \eqref{Br_vec}, \eqref{eq:tau1} -- \eqref{eq:tau3} in Appendix \ref{app:production}).
Therefore the iso-contour for a specific value $N_{detector}$ consists of the values $M_N, |U|^2$ obeying the equation:
\begin{eqnarray}
\label{eq:N_N2}
N_{detector} & = & |U|^2 \mathcal{N}_N \zeta_N  \int_0^{p_{N_L}^{max}}dp_{N_L} f_1(p_{N_L}) \int_0^{l_{decay~pipe}} dz_H f_2(z_H, p_{N_L}) \times\nonumber\\
&& \times exp\left(-\frac{L - z_H}{T_N}\frac{M_N}{p_{N_L}} |U|^2 \right) \left(1 - exp\left(-\frac{\Delta l}{T'_N}\frac{M_N}{p_{N_L}} |U|^2 \right) \right).
\end{eqnarray}

When the value of the sterile neutrino lifetime satisfies $\tau_N \gg (L-z_H)\frac{M_N}{p_{N_L}}$, Eq. \eqref{eq:probability3} has a simple limit:
\begin{equation}
\label{eq:probability4}
P_1\left( L < z_H + l_N < L +\Delta l \right) \approx \frac{\Delta l}{T'_N}\frac{M_N}{p_{N_L}} |U|^2.
\end{equation}
In that case almost none of the produced sterile neutrinos decay before they reach the detector and only the small portion of them decay in the detector volume.
From Eq. \eqref{eq:N_N2} we obtain that $N_{detector} = I \times |U|^4$, where numerical coefficient $I$ doesn't depend on $|U|^2$.
Therefore, for an iso-contour for a value $N_{detector}$, if $|U|^2 \ll \frac{T_N}{L-z_H}\frac{p_{N_L}}{M_N}$, Eq. \eqref{eq:N_N2} can be rewritten in a simple form:
\begin{equation}
\label{eq:lower_limit}
|U|^2 = \sqrt{\frac{N_{detector}}{I}}.
\end{equation}

The second important approximation of Eq. \eqref{eq:probability3} is obtained for $\tau'_N \ll \Delta l\frac{M_N}{p_{N_L}}$:
\begin{equation}
\label{eq:probability5}
P_2\left( L < z_H + l_N < L +\Delta l \right) \approx exp\left(-\frac{L - z_H}{\tau_N}\frac{M_N}{p_{N_L}}\right).
\end{equation}
In that case almost all produced sterile neutrinos decay before they reach the detector and almost all of the sterile neutrinos that reached the detector decay inside the detector volume.
Therefore the iso-contour for $N_{detector}$ consists of the values $M_N, |U|^2$ that satisfy somewhat easier equation:
\begin{eqnarray}
N_{detector} & = & |U|^2 \mathcal{N}_N \zeta_N  \int_0^{p_{N_L}^{max}}dp_{N_L} f_1(p_{N_L}) \times\nonumber\\
&& \times \int_0^{l_{decay~pipe}} dz_H f_2(z_H, p_{N_L}) exp\left(-\frac{L - z_H}{T_N}\frac{M_N}{p_{N_L}} |U|^2 \right).
\end{eqnarray}

For a fixed value of $M_N$, the resulting iso-contours consist of two values of $|U|^2$, which, for the most part of the considered mass range, satisfy the conditions of these two approximations.
Staring with some value of $M_N$ they are no more viable, and one has to use Eq. \eqref{eq:N_N2} as it is.

\section{Results}
\label{sec:results}

\begin{figure}
\begin{center}
\includegraphics[width=0.8\linewidth]{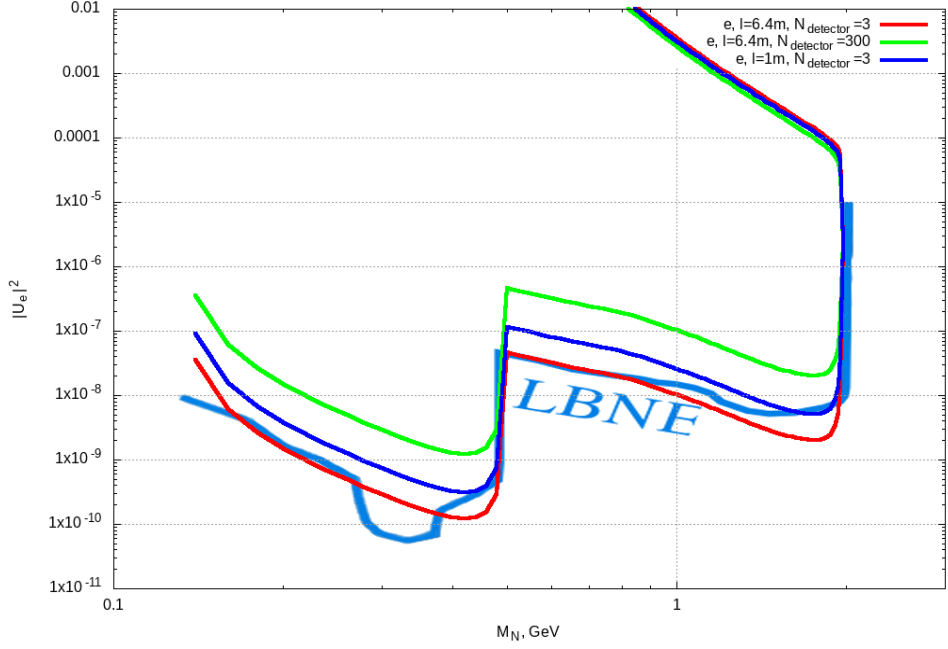}
\caption{The number of expected sterile neutrino decays in the detector volume in the plane $M_N - |U|^2$ for the case of mixing with the electron neutrino. LBNE (steelblue) line is a previous sensitivity estimate \cite{Adams:2013qkq}. Red line is our estimate for $\Delta l= 6.4$m and $N_{detector} = 3$, blue line is for $\Delta l= 1.0$m and $N_{detector} = 3$, and green line is for $\Delta l= 6.4$m and $N_{detector} = 300$.}
\label{fig:sensitivity-e}
\end{center}
\end{figure}

\begin{figure}
\begin{center}
\includegraphics[width=0.8\linewidth]{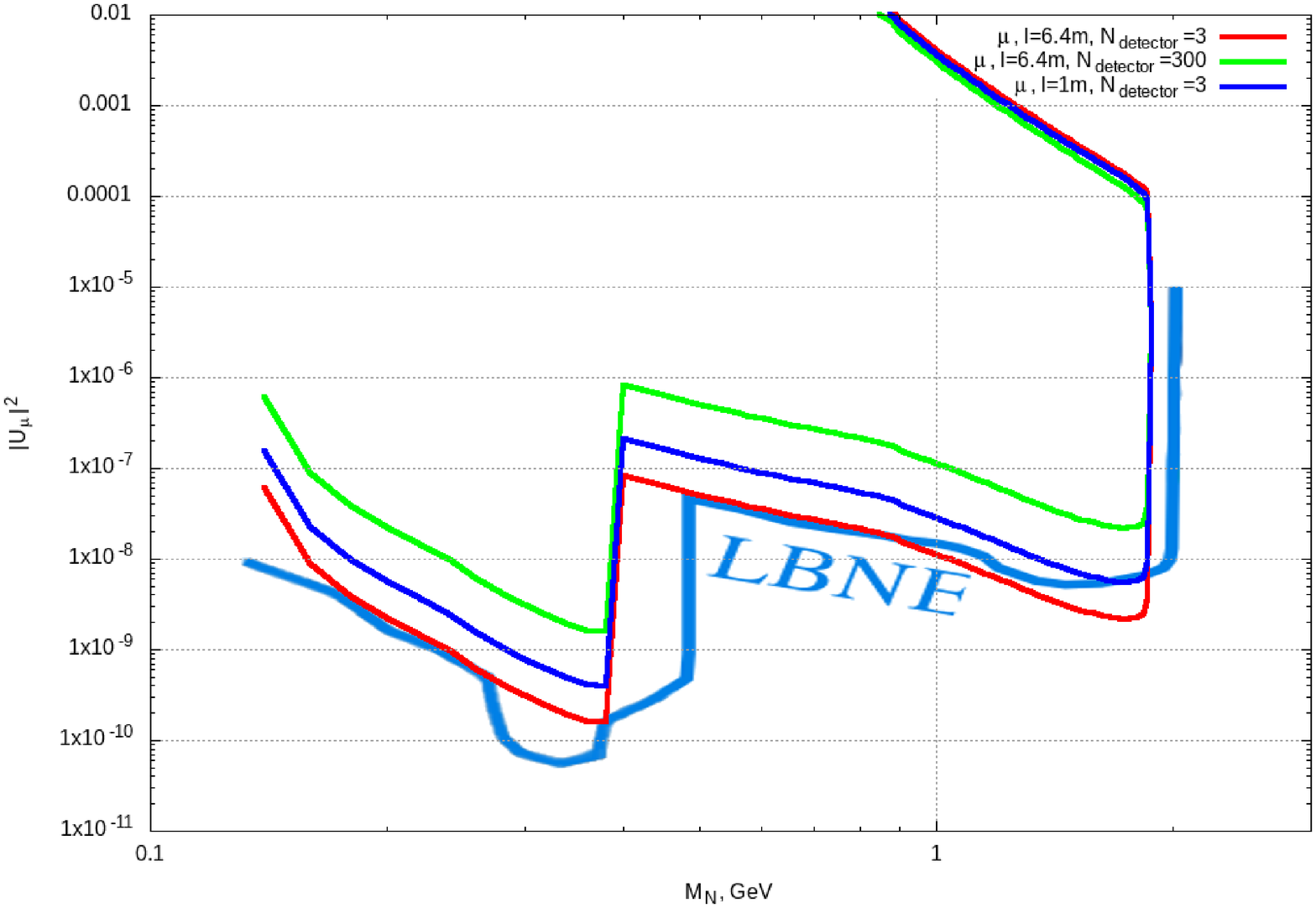}
\caption{The number of expected sterile neutrino decays in the detector volume in the plane $M_N - |U|^2$ for the case of mixing with the muon neutrino. LBNE (steelblue) line is a previous sensitivity estimate \cite{Adams:2013qkq}. Red line is our estimate for $\Delta l= 6.4$m and $N_{detector} = 3$, blue line is for $\Delta l= 1.0$m and $N_{detector} = 3$, and green line is for $\Delta l= 6.4$m and $N_{detector} = 300$.}
\label{fig:sensitivity-mu}
\end{center}
\end{figure}

\begin{figure}
\begin{center}
\includegraphics[width=0.8\linewidth]{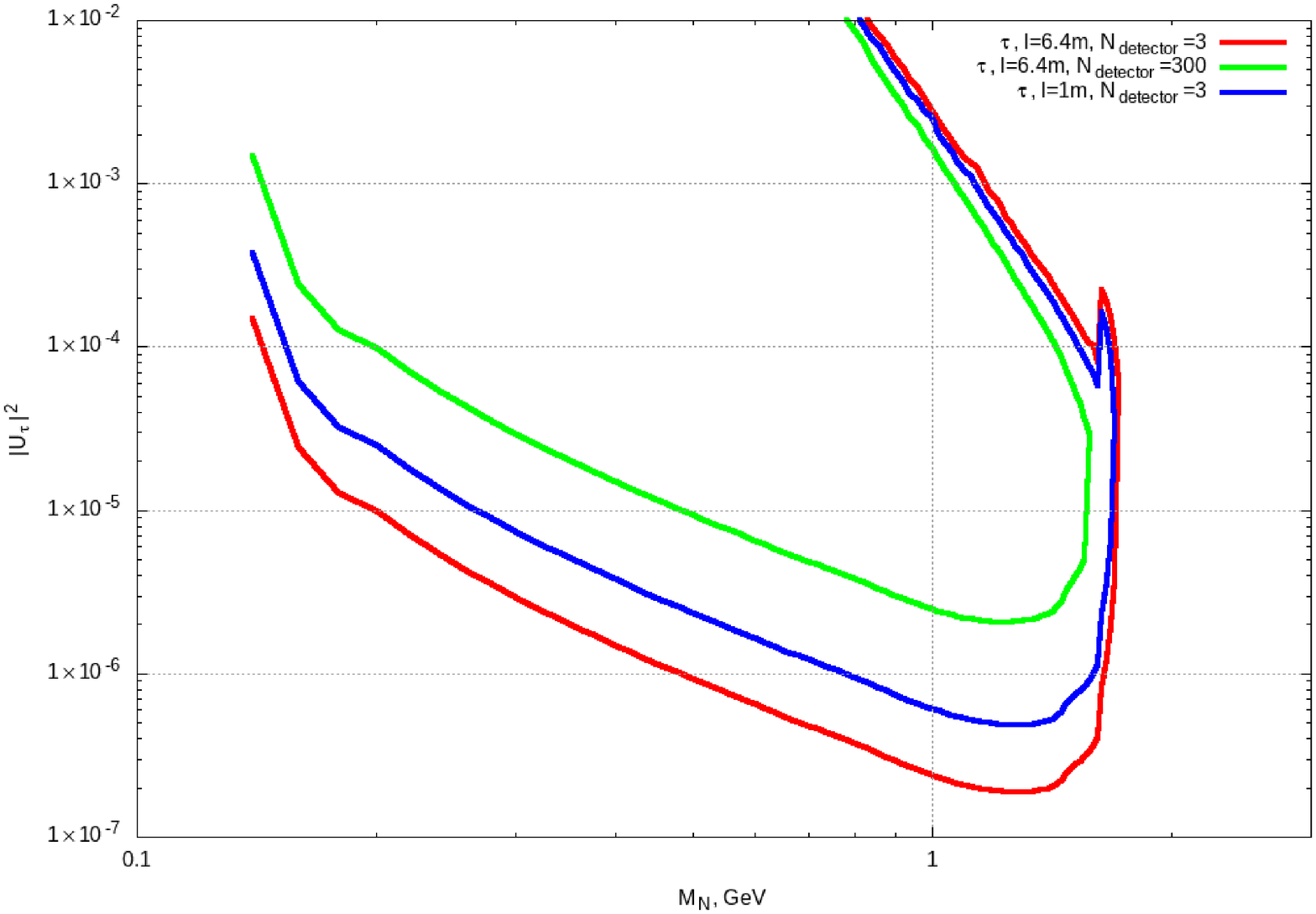}
\caption{The number of expected sterile neutrino decays in the detector volume in the plane $M_N - |U|^2$ for the case of mixing with the tauon neutrino. Red line is our estimate for $\Delta l= 6.4$m and $N_{detector} = 3$, blue line is for $\Delta l= 1.0$m and $N_{detector} = 3$, and green line is for $\Delta l= 6.4$m and $N_{detector} = 300$. The small feature at $M_N=1.63$ GeV is due to the disappearance of the channel $\tau \to \pi N$, that is dominating up to that point.}
\label{fig:sensitivity-tau}
\end{center}
\end{figure}

In this Section, we present our iso-contours and analysis.
We take beam properties and geometrical sizes as described in Ref. \cite{Acciarri:2016crz,Acciarri:2015uup,Strait:2016mof,Acciarri:2016ooe}: $N_{POT} = 1.1 \times 10^{22}$ (this corresponds to the total expected number of protons-on-target over ten years), $\Delta h = 3.5$m, $L=574$ m, $l_{decay\,pipe}=194$ m, $r_{decay\,pipe}=2$ m, see Tabs. \ref{tab:proton}, \ref{tab:geometry}.

For simplicity, in our analysis we vary only two parameters: the detector effective length $\Delta l$ and the number $N_{detector}$ of the sterile neutrino decays inside the detector volume.
In the idealistic situation when there is absolutely no background, according to Poisson distribution $N_{detector} = 3$ should be enough to place the limit at 95$\%$ CL to announce the discovery of the heavy sterile neutrinos.
Unfortunately, the absence of the active neutrino events background isn't a realistic assumption for the DUNE Near Detector.
The main goal of the Near Detector is to characterize the beam of active neutrinos.
Therefore it will be designed in such a way as to increase the probability of neutrino interactions inside the detector.
From the point of view of a sterile neutrino search, such interactions would serve as a background.
We avoid all the issues of the experimental detection efficiencies and reconstruction effects by presenting only the iso-contours for the number of expected heavy neutrino decays into $\it{visible}$ modes inside the detector volume, in the plane $M_N - |U|^2$.

We present our results for mixing with electron, muon and tau neutrino in Figs. \ref{fig:sensitivity-e}, \ref{fig:sensitivity-mu} and \ref{fig:sensitivity-tau} respectively.
For the red lines we take $N_{detector} = 3$ and for green lines we take $N_{detector} = 300$.
For these lines we take the currently considered detector length $\Delta l= 6.4$m from Ref. \cite{Acciarri:2016ooe}.
All iso-contours are calculated in accordance with Eq. \eqref{eq:N_N2}.

There is another possibility we point out.
If we had additional free space in front of the main detector we could place there a small additional detector, sensitive to sterile neutrino decays.
To reduce the active neutrino event background, as well as to minimize the effect of the additional detector on active neutrino study, the additional detector should be almost empty inside.
That would allow for detecting of the sterile neutrino decays in this empty space, where are few active neutrino interactions.
Depending on the design, it could provide better sensitivity to mixing with active neutrinos than the main detector.
In Figs. \ref{fig:sensitivity-e}, \ref{fig:sensitivity-mu}, \ref{fig:sensitivity-tau} we show that case with blue lines, for which we take $\Delta l= 1$m and $N_{detector} = 3$.
From these figures, one can see that, depending on its configuration, a small additional detector with good efficiency could provide better limits than the main detector overburdened with active neutrino background.

For reference we present in Figs. \ref{fig:sensitivity-e}, \ref{fig:sensitivity-mu} the estimate from LBNE design report \cite{Adams:2013qkq}.
That estimate was made simply by rescaling of the CHARM \cite{Bergsma:1985is} and CERN PS191 \cite{Bernardi:1987ek} results, taking into account the relevant proton beam and detector geometry parameters of LBNE and CHARM and PS191 experiments.
The length of the LBNE Near Detector was taken to be $\Delta l = 30$m.
These lines were calculated for the case when sterile neutrino mixes with every type of active neutrinos, while we present the mixing with a specific type.
This results in the difference in the shape of the curve.
Due to the difference in masses between electron and muon, decays of kaons into the sterile neutrino stop at lower masses of the sterile neutrino $M_N$ for the mixing with muon neutrino than for the mixing with electron neutrino.
In Figs. \ref{fig:sensitivity-e}, \ref{fig:sensitivity-mu} one can see these steps at $M_N \sim M_K -m_e$ and $M_N \sim M_K - m_\mu$ respectively.
For LBNE line this shift occurs in two steps, as the muon part of mixing disappears at lower sterile neutrino mass than the electron part.
We didn't find estimates for LBNE limits on mixing with tau neutrino, so in Fig. \ref{fig:sensitivity-tau} we present only our estimates.

One can find current limits on the mixing (and some predictions) in Ref. \cite{Drewes:2018irr,SHiP:2018xqw,Curtin:2018mvb,Alekhin:2015byh,Ariga:2018uku}.

\section{The validity of the results}
\label{sec:background}
In this Section, we estimate the active neutrino flux as a validation of our results and present some ideas on the scale of the signal to the background ratio in the sterile neutrino search and how it can be improved.

\begin{figure}
\begin{center}
\includegraphics[width=0.75\linewidth]{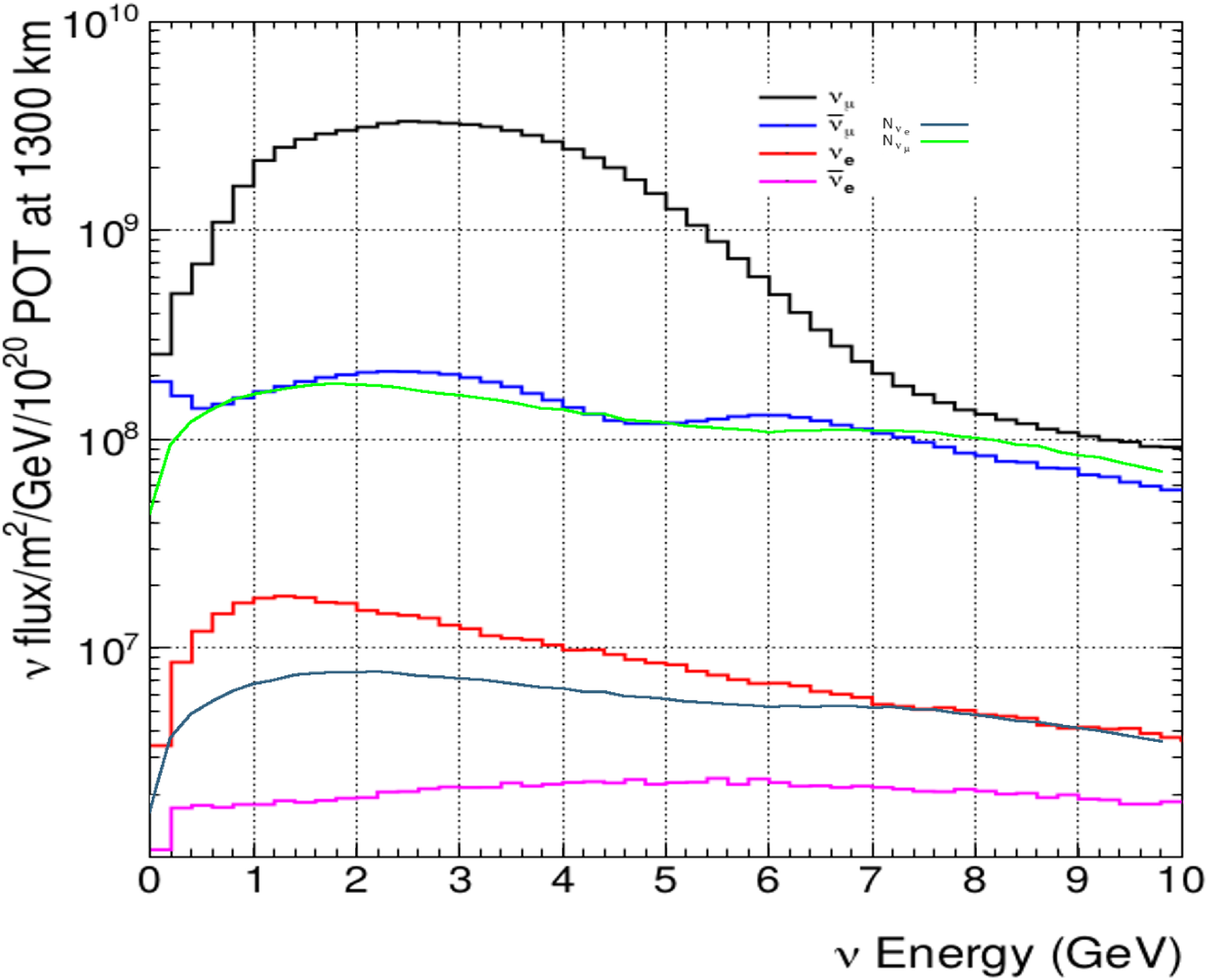}
\caption{Neutrino fluxes adopted from Ref. \cite{Acciarri:2015uup} (black, blue, red and violet lines) for the neutrino mode, generated with a 120 GeV primary proton beam and our corresponding estimate without the horns (steelblue and green lines).}
\label{fig:active1}
\end{center}
\end{figure}

\begin{figure}
\begin{center}
\includegraphics[width=0.75\linewidth]{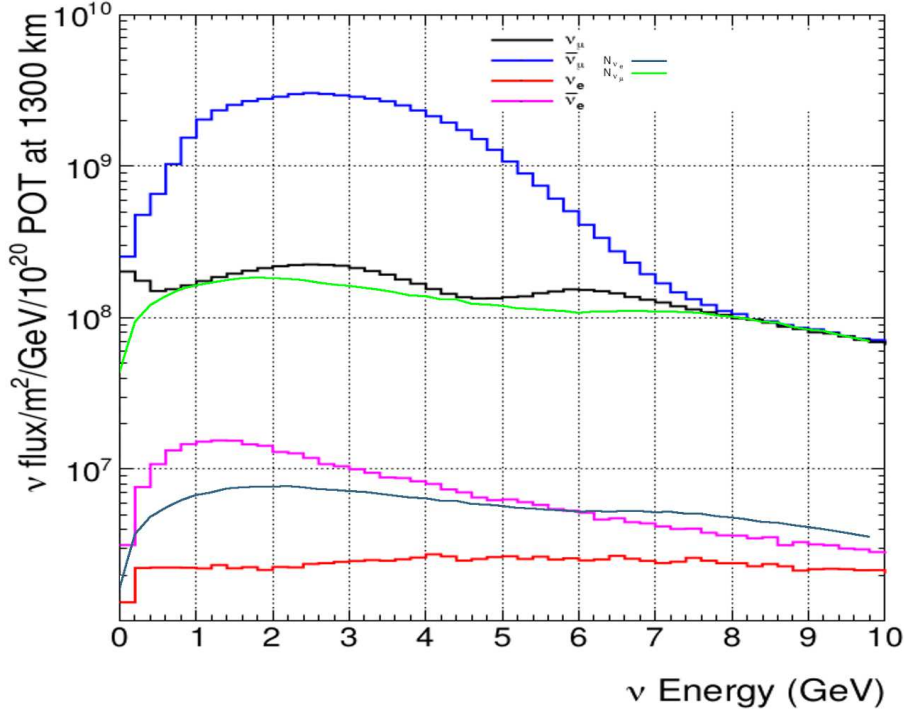}
\caption{Neutrino fluxes adopted from Ref. \cite{Acciarri:2015uup} (black, blue, red and violet lines) for the antineutrino mode, generated with a 120 GeV primary proton beam and our corresponding estimate without the horns (steelblue and green lines).}
\label{fig:active2}
\end{center}
\end{figure}

We have performed a special simulation of the flux of $M_N = 0.01$ eV sterile neutrinos through the Far Detector with $N_{POT} = 10^{20}s$ to check our assumptions of the meson production and decay and compare them with the results from Ref. \cite{Acciarri:2015uup}.
The basic idea is that the obtained results should more or less correspond to active neutrino fluxes from pion and kaon decays.
For that purpose we consider for the ``muon neutrino'' $N_\mu$ the following processes: $\pi^+ \to \mu^+ N_\mu, K^+ \to \mu^+ N_\mu, K^+ \to \pi^0 \mu^+ N_\mu, K^0 \to \pi^- \mu^+ N_\mu$ as well as the decay $\mu^+ \to e^+ N_e N_\mu$ for muons produced in these processes; for the number of the ``electron neutrinos'' $N_e$ we have the same muon decays and the processes  $K^- \to \pi^0 e^- N_e, K^0 \to \pi^+ e^- N_e$.
The fluxes of the neutrino and anti-neutrino are considered to be the same because we don't account for the presence of the horns focusing systems.
Horns focusing systems affect all charged particles, but particularly the
charged pions and kaons, as they are relatively long-lived.
As the name implies, horns would focus these particles, causing more of them to fly in the direction of the detector.
Usually, horns are specialized to focus pions, as their flux is much higher than the flux of kaons.
The usage of horns in other experiments increased the resulting flux of neutrinos by several times.
Recall that, for the sterile neutrino search, active neutrino events serve as a background.
It should be compensated a little because of the focusing of kaons that produce sterile neutrinos, but the overall effect of horns is considered to be negative for the sterile neutrino search.

We present our results, as well as the neutrino fluxes adopted from Ref. \cite{Acciarri:2015uup} in Figs. \ref{fig:active1}, \ref{fig:active2}.
As we don't account for the presence of the horns focusing systems, our results should roughly correlate to the anti-neutrino flux in the neutrino mode (and neutrino flux in anti-neutrino mode), or be a bit higher.
This behaviour can be seen in  Figs. \ref{fig:active1}, \ref{fig:active2}.
Note that neutrinos with higher energy are less affected by horns systems and results from Ref. \cite{Acciarri:2015uup} for this part of the spectrum are in good correlation with our estimate.
We note that the shape of the resulting energy-flux curve is somewhat different from Ref. \cite{Acciarri:2015uup}.
Besides the presence of the horns that can be affected by more rare processes that aren't accounted for in our estimate.
Overall, we find these results satisfactory given the crudeness of the estimate and the factor of the presence of the horns.

There are several methods that are used to improve the signal to the background ratio.
One such way is to take into consideration only certain decay modes.
Among various sterile neutrino decay modes, the most promising ones are the two body decays.
The products of two body decays have a fixed momentum in the decaying particle rest frame.
Its value depends only on the decaying particle mass.
This fact allows for a more precise reconstruction of the mass of decaying particle than in the case of three-or-more body decays.

Let's consider a specific example: the decay $N \to \pi^+ \mu^-$.
The Monte-Carlo simulation results for $N_{POT}= 10^{20}$ presented in Ref. \cite{Acciarri:2015uup} suggest 44 000 events of the $\nu_\mu X \to \pi^+ \mu^- X$ type, where $X$ is the atom of the target.
Hence for $N_{POT}= 1.1 \times 10^{22}$, adopted in this paper, the number of background events integrated over the energy easily reaches $4.8 \times 10^6$!
For the sterile neutrino two body decay, however, not all of these events serve as a background.
If one reconstructs the energy distribution of the detected $\mu, \pi$,
the products of the sterile neutrino decays would form characteristic peaks.
Only events for $\mu, \pi$ with the same energy as these peaks would serve as a background for the sterile neutrino signal.
Obviously similar considerations apply to other decay modes as well.
Unfortunately, the simulation/reconstruction of such energy distribution depends heavily on the detector specifics and is rather hard to perform.
A detailed study of detector efficiency will be possible only after the final decision on the design of the Near Detector is made.

An additional possibility is to account for the fact that sterile neutrinos that reach the detector have very small transversal momentum.
Therefore the sum of the resulting particles transversal momentum should also be close to zero.
If one cuts all events for which it doesn't hold true, one would reduce the amount of background.

\section{Conclusions}
\label{sec:conclusions}
In this paper, we calculated the prospects for the sterile neutrino search in the upcoming experiment DUNE.
We present the iso-contours for the number of expected heavy neutrino decays inside the detector volume, in the plane $M_N - |U|^2$.
Our more optimistic predictions are approximately of the same order as the previous estimates, while more conservative ones lay higher than it, but still lower than the current limits.
We point out that for the search of sterile neutrinos an additional small detector, that is almost empty inside, could provide better sensitivity than the main detector.
Another possibility to enhance the signal to background ratio can be a special run with a really short proton spill duration and lower proton beam energy.
Overall our estimates show that while DUNE main scientific goal is the measurement of active neutrino parameters, it would still be able to probe currently unrestricted part of the sterile neutrino parameter space.

\section*{Acknowledgments}
We would like to thank D. Gorbunov and Y. Kudenko for valuable discussions. The work was supported by the RSF grant 17-12-01547.
\appendix
\section{Parameters}
\label{sec:parameters}
In this Section, we list the experimental values of the parameters used in our work.

Lepton masses $l \in \{ e, \mu, \tau\}$ :
$M_e = 0.5109989461$ MeV, $M_\mu = 105.6583745$ MeV, $M_\tau = 1.77686$ GeV \cite{Tanabashi:2018oca}.

Tauon average lifetime $\tau_\tau = 2.903 \times 10^{-4}$ ns \cite{Tanabashi:2018oca}.

Fermi constant: $G_F = 1.16637877\times 10^{-5}$ GeV$^{-2}$ \cite{Tanabashi:2018oca}.

Weinberg angle: $\sin^2 \theta_W = 0.23122$ \cite{Tanabashi:2018oca}.

$\rho$-meson decay constant: $g_\rho=0.162$ GeV$^2$ \cite{Bondarenko:2018ptm}.
\begin{table}[ht] 
\label{tab:ckm}
\begin{center}
\begin{tabular}{|*{7}{c|}}
\hline
$V_{ud}$ & $V_{us}$ & $V_{cd}$ & $V_{cs}$ & $ V_{ub}$ & $V_{cb}$\\
\hline
0.97420 & 0.2243 & 0.218 & 0.997 & 0.00394 & 0.0422\\
\hline
\end{tabular}
\end{center}
\caption{CKM-matrix elements \cite{Tanabashi:2018oca}.}
\end{table}

\begin{table}[ht]
\begin{center}
\begin{tabular}{|l|*{4}{c|}}
\hline
$H$ & $M_H$, MeV \cite{Tanabashi:2018oca} & $\tau_H$, ns \cite{Tanabashi:2018oca} & $f_H,$ MeV \cite{Bondarenko:2018ptm}\\
\hline
$\pi^+$ & 139.57061 & 26.033 & 130.2\\
\hline
$\pi^0$ & 134.977 & 8.52e-8 & 130.2\\
\hline
$K^+$ & 493.677 & 12.38 & 155.6\\
\hline
$K^0_L$ & 497.611 & 51.16 & \\
\hline
$K^0_S$ & 497.611 & 0.089564 & \\
\hline
$\eta$ & 547.862 & & 81.7\\
\hline
$\rho$ & 775.26 & & \\
\hline
$\eta '$ & 957.78 & & -94.7\\
\hline
$D^+$ & 1869.65 & 1.04e-3 & 212\\
\hline
$D^0$ & 1864.83 & 4.101e-4 & \\
\hline
$D_s^+$ & 1968.34 & 5.04e-4 & 249\\
\hline
$B^+$ & 5279.32 & 1.638e-3 & 187\\
\hline
$B^0$ & 5279.63 & 1.52e-3 & \\
\hline
$B_s^0$ & 5366.89 & 1.509e-3 & \\
\hline
$B_c^+$ & 6274.9 & 5.07e-4 & 434\\
\hline
\end{tabular}
\end{center}
\caption{Relevant meson decay parameters \cite{Tanabashi:2018oca,Bondarenko:2018ptm}.}
\label{tab:meson}
\end{table}

\subsection{Form-factors}
Basic formula \cite{Bondarenko:2018ptm}:
\begin{equation}
f(q^2) = \frac{1}{1- q^2/M_{pole}^2} \sum_{n=0}^{N-1} a_n \left[ \left(z(q^2)\right)^n - (-1)^{n-N} \frac{n}{N} \left(z(q^2)\right)^N\right],
\end{equation}
where
\begin{equation}
z(q^2) \equiv \frac{\sqrt{t_+-q^2}-\sqrt{t_+-t_0}}{\sqrt{t_+-q^2}+\sqrt{t_+-t_0}},
\end{equation}
with
\begin{equation}
t_+ = \left( m_H + m_{H'} \right)^2,
\end{equation}
\begin{equation}
t_0 = \left( m_H + m_{H'} \right) \left( \sqrt{m_H} - \sqrt{m_{H'}} \right)^2.
\end{equation}
\subsubsection{K meson form factors}
Form factors of $K\to\pi$ transition are well described by the linear approximation:
\begin{equation}
f^{K\pi}_{+,0}(q^2) = f^{K\pi}_{+,0}(0) \left( 1+ \lambda_{+,0} \frac{q^2}{m^2_{\pi^+}}\right).
\end{equation}

\begin{table}[ht] 
\label{tab:form-factor-K}
\begin{center}
\begin{tabular}{|l|*{3}{c|}}
\hline
$H, H'$ & $f_{+,0}(0)$ & $\lambda_+$ & $\lambda_0$\\
\hline
$K^0, \pi^+$ & 0.970 & 0.0267 & 0.0117\\
\hline
$K^+, \pi^0$ & 0.970 & 0.0277 & 0.0183\\
\hline
\end{tabular}
\end{center}
\caption{Best fit parameters \cite{Bondarenko:2018ptm} for the form factors of the $K\to\pi$ transition.}
\end{table}

\subsubsection{D meson form factors}
Form factors of $D \to K,\pi$:
\begin{equation}
f(q^2) = \frac{f(0)-c \left( z(q^2) - z(0) \right)\left( 1 + \frac{z(q^2)+z(0)}{2} \right)}{1 - P q^2}.
\end{equation}

\begin{table}[ht] 
\label{tab:form-factor-D}
\begin{center}
\begin{tabular}{|*{4}{c|}}
\hline
$f$ & $f(0)$ & c & $P($GeV$^-2)$\\
\hline
$f_+^{DK}$ & 0.7647 & 0.066 & 0.224\\
\hline
$f_0^{DK}$ & 0.7647 & 2.084 & 0\\
\hline
$f_+^{D\pi}$ & 0.6117 & 1.985 & 0.1314\\
\hline
$f_0^{D\pi}$ & 0.6117 & 1.188 & 0.0342\\
\hline
\end{tabular}
\end{center}
\caption{Best fit parameters \cite{Bondarenko:2018ptm} for the form factors of the $D \to K,\pi$ transition.}
\end{table}

Form factors of $D \to \eta$:
\begin{equation}
f^{D_s\eta}_+(q^2) = \frac{f^{D_s\eta}_+(0)}{\left(1 - \frac{q^2}{M_{D_s^*}^2}\right) \left(1 - \alpha^{D_s\eta}_+ \frac{q^2}{M_{D_s^*}^2}\right)},
\end{equation}
where $f^{D_s\eta}_+(0)=0.495, M_{D_s^*} = 2.112, \alpha^{D_s\eta}_+ = 0.198$ \cite{Bondarenko:2018ptm}.
\begin{equation}
f^{D_s\eta}_0(q^2) = \frac{f^{D_s\eta}_0(0)}{1 - \alpha^{D_s\eta}_0 \frac{q^2}{M_{D_s^*}^2}},
\end{equation}
$f^{D_s\eta}_0(q^2)$ is not well constrained by experimental data, so we take $f^{D_s\eta}_0(0) = f^{D_s\eta}_+(0)$ and $\alpha^{D_s\eta}_0 = 0$ \cite{Bondarenko:2018ptm}.
\subsubsection{B meson form factors}
Form factors of $B \to D,\pi$:

\begin{table}[ht] 
\label{tab:form-factor-B}
\begin{center}
\begin{tabular}{|*{5}{c|}}
\hline
$f$ & $M_{pole}$ GeV & $a_0$ & $a_1$ & $a_2$\\
\hline
$f_+^{B_{(s)}D_{(s)}}$ & $\infty$ & 0.909 & -7.11 & 66\\
\hline
$f_0^{B_{(s)}D_{(s)}}$ & $\infty$ & 0.794 & -2.45 & 33\\
\hline
$f_+^{BK}$ & 5.325 & 0.360 & -0.828 & 1.1\\
\hline
$f_0^{BK}$ & 5.65 & 0.233 & 0.197 & 0.18\\
\hline
$f_+^{B\pi}$ & 5.325 & 0.404 & -0.68 & -0.86\\
\hline
$f_0^{B\pi}$ & 5.65 & 0.490 & -1.61 & 0.93\\
\hline
\end{tabular}
\end{center}
\caption{Best fit parameters \cite{Bondarenko:2018ptm} for the form factors of the $B \to D,\pi$ transition.}
\end{table}
\subsubsection{Meson form factors for decay into vector meson}
Standard axial form factors $A_0(q^2), A_1(q^2), A_2(q^2)$ and vector form factor $V(q^2)$ can be parameterized as:
\begin{eqnarray}
V(q^2) & = & \frac{f^{hh'}_{V}}{\left(1- \frac{q^2}{M_V^{h^2}}\right)\left(1 - \sigma^{hh'}_{V} \frac{q^2}{M_V^{h^2}} - \zeta_{V}^{hh'} \frac{q^4}{M_V^{h^4}}\right)},\\
A_0(q^2) & = & \frac{f^{hh'}_{A_0}}{\left(1- \frac{q^2}{M_P^{h^2}}\right)\left(1 - \sigma^{hh'}_{A_{0}} \frac{q^2}{M_V^{h^2}} - \zeta_{A_{0}}^{hh'} \frac{q^4}{M_V^{h^4}}\right)},\\
A_{1,2}(q^2) & = & \frac{f^{hh'}_{A_{1,2}}}{1 - \sigma^{hh'}_{A_{1,2}} \frac{q^2}{M_V^{h^2}} - \zeta_{A_{1,2}}^{hh'} \frac{q^4}{M_V^{h^4}}}.
\end{eqnarray}
\begin{table}[ht] 
\label{tab:form-factor-vec}
\begin{center}
\begin{tabular}{|*{6}{c|}}
\hline
$hh'$ & $DK^*$ & $BD^*$ & $B\rho$ & $B_sD^*$ & $B_sK$\\
\hline
$f^{hh'}_{V}$ & 1.03 & 0.76 & 0.295 & 0.95 & 0.291\\
\hline
$f^{hh'}_{A_0}$ & 0.76 & 0.69 & 0.231 & 0.67 & 0.289\\
\hline
$f^{hh'}_{A_1}$ & 0.66 & 0.66 & 0.269 & 0.70 & 0.287\\
\hline
$f^{hh'}_{A_2}$ & 0.49 & 0.62 & 0.282 & 0.75 & 0.286\\
\hline
$\sigma^{hh'}_{V}$ & 0.27 & 0.57 & 0.875 & 0.372 & -0.516\\
\hline
$\sigma^{hh'}_{A_0}$ & 0.17 & 0.59 & 0.796 & 0.350 & -0.383\\
\hline
$\sigma^{hh'}_{A_1}$ & 0.3 & 0.78 & 0.54 & 0.463 & 0\\
\hline
$\sigma^{hh'}_{A_2}$ & 0.67 & 1.4 & 1.34 & 1.04 & 1.05\\
\hline
$\zeta^{hh'}_{V}$ & 0 & 0 & 0 & 0.561 & 2.10\\
\hline
$\zeta^{hh'}_{A_0}$ & 0 & 0 & 0.055 & 0.600 & 1.58\\
\hline
$\zeta^{hh'}_{A_1}$ & 0.2 & 0 & 0 & 0.510 & 1.06\\
\hline
$\zeta^{hh'}_{A_2}$ & 0.16 & 0.41 & -0.21 & 0.070 & -0.074\\
\hline
$M^{h}_{P}$ (GeV)& 1.969 & 6.275 & 5.279 & 6.275 & 5.367\\
\hline
$M^{h}_{V}$ (GeV)& 2.112 & 6.331 & 5.325 & 6.331 & 5.415\\
\hline
\end{tabular}
\end{center}
\caption{Best fit parameters \cite{Bondarenko:2018ptm} of the meson form factors of the decays into vector meson.}
\end{table}

\section{Formulae}
\label{sec:formulae}
\subsection{Sterile neutrino decays}
\label{app:decay}
2-particle sterile neutrino decays \cite{Gorbunov:2007ak,Bondarenko:2018ptm}:
\begin{eqnarray}
\label{eq:decay_first}
\Gamma (N \to H^0\nu_\alpha) & = & \frac{|U_\alpha|^2}{32 \pi} G_F^2 f_{H^0}^2 M_N^3 \left( 1 - \frac{M_{H^0}^2}{M_N^2} \right)^2,\\
\label{eq:decay_second}
\Gamma (N \to H^+ l^-_\alpha) & = & \frac{|U_\alpha|^2}{16 \pi} G_F^2 |V_H|^2 f_H^2 M_N^3 \left( \left( 1 - \frac{M_l^2}{M_N^2} \right)^2 - \frac{M_H^2}{M_N^2} \left( 1+ \frac{M_l^2}{M_N^2}\right) \right) \times\nonumber\\
&&\times \sqrt{\left( 1- \frac{(M_H-M_l)^2}{M_N^2}\right) \left(( 1- \frac{(M_H+M_l)^2}{M_N^2} \right)},\\
\Gamma (N \to V^+ l^-_\alpha) & = & \frac{|U_\alpha|^2}{16 \pi} \frac{g^2_{V^+}}{M^2_{V^+}} G_F^2 |V_{V}|^2 M_N^3  \times \nonumber\\
&& \times \left( \left( 1 - \frac{M_l^2}{M_N^2} \right)^2 + \frac{M^2_{V^+}}{M_N^2} \left( 1 + \frac{M_l^2 - 2 M^2_{V^+}}{M_N^2}\right) \right) \times\nonumber\\
&&\times \sqrt{\left( 1- \frac{(M_{V^+} -M_l)^2}{M_N^2}\right) \left(( 1- \frac{(M_{V^+} +M_l)^2}{M_N^2} \right)},\\
\Gamma (N \to V^0 \nu_\alpha) & = & \frac{|U_\alpha|^2}{32 \pi} \frac{g^2_{V^0} k_V^2}{M^2_{V^0}} G_F^2 M_N^3 \left( 1 +2 \frac{M^2_{V^0}}{M_N^2} \right) \left( 1 - \frac{M_{V^0}^2}{M_N^2} \right)^2,
\end{eqnarray}
where $H^0 \in \{ \pi^0, \eta , \eta', ...\}, H^+ \in \{ \pi^+, K^+ , D^+, ...\}, V^0 \in \{\rho^0, ...\}, V^+ \in \{\rho^+, ...\}$ (see Tab. \ref{tab:meson}).
In this work we use only $k_\rho  =  1 - 2 \sin^2 \theta_W$.

3 particle sterile neutrino decays \cite{Gorbunov:2007ak,Bondarenko:2018ptm}:
\begin{eqnarray}
\label{eq:decay_3nu}
\Gamma \left( N \to \nu_\alpha \sum_\beta \bar{\nu}_\beta \nu_\beta \right) & = & \frac{G_F^2 M_N^5}{192 \pi^3} |U_\alpha |^2 ,\\
\Gamma \left( N \to l^-_{\alpha \neq \beta} l^+_\beta \nu_\beta \right) & = & \frac{G_F^2 M_N^5}{192 \pi^3} |U_\alpha |^2 \left( 1 - 8 x_l^2 + 8 x_l^6 - x_l^8 -12 x_l^4 \log x_l^2 \right),\\
&& x_l = \frac{\max[M_{l_\alpha}, M_{l_\beta}]}{M_N},\nonumber\\
\Gamma \left( N \to \nu_\alpha l^-_\beta l^+_\beta \right) & = & \frac{G_F^2 M_N^5}{192 \pi^3} |U_\alpha |^2 \Big[ \left( C_1 (1-\delta_{\alpha \beta}) + C_3 \delta_{\alpha \beta} \right) \times\nonumber\\
&&\times \left( (1 - 14 x_l^2 - 2 x_l^4 - 12 x_l^6) \sqrt{1-4 x_l^2} +12 x_l^4 (x_l^4 -1) L \right) +\nonumber\\
&& + 4 \left( C_2 (1-\delta_{\alpha \beta}) + C_4 \delta_{\alpha \beta} \right) \times \Big(  x_l^2 (2 + 10 x_l^2 - \nonumber\\
\label{eq:decay_last}
&& - 12 x_l^4) \sqrt{1-4 x_l^2} + 6 x_l^4 (1 - 2 x_l^2 + 2 x_l^4) L \Big) \Big],
\end{eqnarray}
where
\begin{displaymath}
L=\log \left[ \frac{1- 3 x_l^2 - (1 - x_l^2) \sqrt{1-4 x_l^2}}{x_l^2 (1 + \sqrt{1-4 x_l^2})}\right], x_l \equiv \frac{M_l}{M_N},
\end{displaymath}
and 
\begin{displaymath}
\begin{array}{rclrcl}
C_1 & = & \frac{1}{4} (1 - 4 \sin^2 \theta_w + 8 \sin^4 \theta_w), & C_2 & = & \frac{1}{2} \sin^2 \theta_w  (2 \sin^2 \theta_w - 1),\\
\nonumber\\
C_3 & = & \frac{1}{4} (1 + 4 \sin^2 \theta_w + 8 \sin^4 \theta_w), & C_4 & = & \frac{1}{2} \sin^2 \theta_w  (2 \sin^2 \theta_w + 1).\\
\end{array}
\end{displaymath}

\subsection{Sterile neutrino production}
\label{app:production}
2-body meson decays with sterile neutrino production \cite{Gorbunov:2007ak}:
\begin{eqnarray}
\label{Br2}
\frac{d Br(H^+ \to l^+_\alpha N)}{d E_N} & = & \tau_H \frac{G_F^2 M_N^2 M_H f_H^2}{8 \pi}  |V_H|^2 |U_\alpha |^2 \Big(1-\frac{M_N^2}{M_H^2}+2\frac{m_l^2}{M_H^2}+\frac{m_l^2}{M_N^2} \times\nonumber\\ 
&&\times \left(1-\frac{m_l^2}{M_H^2}\right)\Big) \sqrt{\left(1+\frac{M_N^2}{M_H^2}-\frac{m_l^2}{M_H^2}\right)^2- 4 \frac{M_N^2}{M_H^2}} \times\nonumber\\ 
&&\times \delta \left(E_N-\frac{M_H^2 - M_l^2 + M_N^2}{2 M_H}\right).
\end{eqnarray}

We take the following decays into two particles: $K^+ \to  l^+ N, D^+ \to  l^+ N, D_s^+ \to  l^+ N, B^+ \to  l^+ N, B_c^+ \to  l^+ N$.

3-particle scalar meson decays with sterile neutrino production \cite{Gorbunov:2007ak}:
\begin{eqnarray}
\label{Br3}
\frac{d Br(H \to H' l_\alpha N)}{d E_N} & = & \tau_H |U_\alpha |^2 C_K^2 \frac{G_F^2 |V_{HH'}|^2}{64 \pi^3 M_H^2} \int_{q^2_{min}}^{q^2_{max}} dq^2\times\nonumber\\ 
&&\times \Big(f_-^2(q^2) \big(q^2 (M_N^2+M_l^2)-(M_N^2-M_l^2)^2\big) +\nonumber\\
&& + 2 f_-(q^2) f_+(q^2) \big(M_N^2 (2 M_H^2 - 2 M_{H'}^2 - 4 E_N M_H - M_l^2 + M_N^2 + q^2) +\nonumber\\
&& \quad+  M_l^2 (4 E_N M_H + M_l^2 - M_N^2 - q^2)\big) +\nonumber\\
&& + f_+^2(q^2) \big((4 E_N M_H + M_l^2 - M_N^2 - q^2) \times \nonumber\\
&& \quad\times (2 M_H^2 - 2 M_{H'}^2 - 4 E_N M_H - M_l^2 + M_N^2 + q^2) -\nonumber\\
&& \quad\quad-  (2 M_H^2 + 2 M_{H'}^2 - q^2) (q^2 - M_N^2 -M_l^2)\big)\Big),
\end{eqnarray}
where $C_K=\frac{1}{\sqrt{2}}$ for $H'=\pi^0$, $C_K=1$ for all other cases \cite{Bondarenko:2018ptm} and $q^2$ range is \cite{Tanabashi:2018oca}:
\begin{eqnarray}
q^2_{min}&=&(E_2^* + E_3^*)^2-\left(\sqrt{E_2^{*2} - M_l^2}+\sqrt{E_3^{*2} - M_N^2}\right)^2,\\
q^2_{max}&=&(E_2^* + E_3^*)^2-\left(\sqrt{E_2^{*2} - M_l^2}-\sqrt{E_3^{*2} - M_N^2}\right)^2,\\
E_2^*&=&\frac{M_H^2 + M_N^2 + M_l^2 - M_{H'}^2 - 2 M_H E_N}{2 m_{12}},\\
E_3^*&=&\frac{M_H E_N - M_N^2}{m_{12}},\\
m_{12}&=&\sqrt{M_H^2 + M_N^2 - 2 M_H E_N},
\end{eqnarray}
$q^2$ range depends on $E_N$.
\cite{Tanabashi:2018oca} provides us with $E_N$ range: $(M_{H'}+M_l)^2 \leq m_{12}^2 \leq (M_H-M_N)^2$.
It is equivalent to $M_N \leq E_N \leq \frac{1}{2 M_H} \left(M_H^2 + M_N^2 -(M_{H'}+M_l)^2\right)$.

3-particle vector meson decays with sterile neutrino production \cite{Gorbunov:2007ak}:
\begin{eqnarray}
\label{Br_vec}
\frac{d Br(H \to V l_\alpha N)}{d E_N} & = & \tau_H |U_\alpha |^2 C_K^2 \frac{G_F^2 |V_{HV}|^2}{32 \pi^3 M_H^2} \int_{q^2_{min}}^{q^2_{max}} dq^2\times\nonumber\\ 
&&\times \Big( \frac{f_2^2(q^2)}{2} \Big( q^2-M_N^2-M_l^2+\omega^2 \frac{\Omega^2- \omega^2}{M_V^2}\Big)+\nonumber\\ 
&& + \frac{f_5^2(q^2)}{2} (M_N^2 + M_l^2) (q^2 - M_N^2 + M_l^2) \Big( \frac{\Omega^4}{4 M_V^2} - q^2\Big)+\nonumber\\
&&+ 2 f_3^2(q^2) M_V^2 \Big( \frac{\Omega^4}{4 M_V^2} - q^2\Big) \Big(M_N^2+M_l^2 -q^2 +\omega^2 \frac{\Omega^2- \omega^2}{M_V^2}\Big)+\nonumber\\
&&+ 2 f_3(q^2) f_5(q^2) (M_N^2 \omega^2 + (\Omega^2- \omega^2) M_l^2)\Big( \frac{\Omega^4}{4 M_V^2} - q^2\Big)+\nonumber\\
&&+ 2 f_1(q^2) f_2(q^2) \big(q^2 (2 \omega^2 - \Omega^2)+\Omega^2(M_N^2-M_l^2)\big)+\nonumber\\
&&+ \frac{f_2(q^2) f_5(q^2)}{2} \Big( \frac{\omega^2 \Omega^2}{M_V^2}(M_N^2 - M_l^2) + \frac{\Omega^4}{M_V^2} M_l^2 +\nonumber\\
&&\quad\quad\quad\quad\quad\quad + 2 (M_N^2 - M_l^2)^2 - 2 q^2 (M_N^2 + M_l^2)\Big)+\nonumber\\
&&+ f_2(q^2) f_3(q^2) \Big( \Omega^2 \omega^2 \frac{ \Omega^2 - \omega^2}{M_V^2}+ 2 \omega^2 (M_l^2-M_N^2) + \Omega^2 (M_N^2 - M_l^2 -q^2) \Big)+\nonumber\\
&&+ f_1^2(q^2) \Big( \Omega^4 (q^2-M_N^2+M_l^2)- 2 M_V^2 (q^4 - (M_N^2-M_l^2)^2)+\nonumber\\
&&\quad\quad\quad\quad\quad\quad +2 \omega^2 \Omega^2 (M_N^2-q^2-M_l^2) + 2  \omega^4 q^2 \Big)\Big),
\end{eqnarray}
where $\omega^2 = M_H^2 - M_V^2 + M_N^2 - M_l^2 - 2 M_H E_N$ and $\Omega^2 = M_H^2 - M_V^2 -q^2$, $C_K=\frac{1}{\sqrt{2}}$ for $H'=\rho^0$, $C_K=1$ for all other cases \cite{Bondarenko:2018ptm}; form factors $f_i(q^2)$ can be expressed via standard axial form factors $A_0(q^2), A_1(q^2), A_2(q^2)$ and vector form factor $V(q^2)$ as:
\begin{equation}
\begin{array}{c}
f_1(q^2) = \frac{V(q^2)}{M_H+M_V},\quad f_2(q^2) = (M_H+M_V) A_1(q^2),\quad f_3(q^2) = - \frac{A_2(q^2)}{M_H+M_V},\\
f_4(q^2) = \frac{1}{q^2} \big(M_V(2 A_0 - A_1 - A_2) + M_H (A_2 - A_1)\big),\quad f_5(q^2) = f_3(q^2) + f_4(q^2).
\end{array}
\end{equation}
\begin{table}[ht] 
\label{tab:2meson-ckm}
\begin{center}
\begin{tabular}{|l|*{7}{c|}}
\hline
& $\pi^+ \to l^+ N$ & $K^+ \to l^+ N$ & $D^+ \to l^+ N$ & $D_s^+ \to l^+ N$ & $B^+ \to l^+ N$ & $B_c^+ \to l^+ N$\\
\hline
$V_H$ & $V_{ud}$ & $V_{us}$ & $V_{cd}$ & $V_{cs}$ & $ V_{ub}$ & $V_{cb}$\\
\hline
\end{tabular}
\end{center}
\caption{2 particle meson decay CKM-matrix elements.}
\end{table}

We take following decays into three particles: $K^0_L \to \pi^+ l^- N,\;K^0_S \to \pi^+ l^- N,\;K^+ \to \pi^0 l^+ N,\;D^0 \to K^+ l^- N,\;D^+ \to K^0 l^+ N,\;D^0 \to \pi^+ l^- N,\;D^+ \to \pi^0 l^+ N,\;D_s^+ \to \eta^0 l^+ N,\;B^+ \to D^0 l^+ N,\;B^0 \to D^+ l^- N,\;B^+ \to \pi^0 l^+ N,\;B^0 \to \pi^+ l^- N,\;D^0 \to K^{+*} l^- N,\;D^+ \to K^{0*} l^+ N,\;B^0 \to D^{+*} l^- N,\;B^+ \to D^{0*} l^+ N,\;B^+ \to \rho^0 l^+ N,\;B^0 \to \rho^+ l^- N,\;B_s^0 \to D_s^+ l^- N,\;B_s^0 \to K^+ l^- N,\;B_s^0 \to D_s^{*+} l^- N,\;B_s^0 \to K^{*+} l^- N$.
\begin{table}[ht] 
\label{tab:3meson-ckm}
\begin{center}
\begin{tabular}{|l|*{6}{c|}}
\hline
& $K^0 \to \pi^+ l^- N$, & $D^0 \to K^+ l^- N$, & $D^0 \to \pi^+ l^- N$, & $B^0 \to D^+ l^- N$, & $B^0 \to \pi^+ l^- N$,\\
& $K^+ \to \pi^0 l^+ N$ & $D^+ \to K^0 l^+ N$,& $D^+ \to \pi^0 l^+ N$ & $B^+ \to D^0 l^+ N$, & $B^+ \to \pi^0 l^+ N$,\\
&& $D^0 \to K^{+*} l^- N$, & $D_s^+ \to \eta^0 l^+ N$ & $B^0 \to D^{+*} l^- N$, & $B^+ \to \rho^0 l^+ N$,\\
&& $D^+ \to K^{0*} l^+ N$ && $B^+ \to D^{0*} l^+ N$ & $B^0 \to \rho^+ l^- N$,\\
&&&& $B_s^0 \to D_s^+ l^- N$ & $B_s^0 \to K^+ l^- N$,\\
&&&& $B_s^0 \to D_s^{*+} l^- N$ & $B_s^0 \to K^{*+} l^- N$\\
\hline
$V_{HH'}$ & $V_{us}$ & $V_{cs}$ & $V_{cd}$ & $V_{cb}$ & $V_{ub}$\\
\hline
\end{tabular}
\end{center}
\caption{3 particle meson decay CKM-matrix elements.}
\end{table}

We consider three cases: mixing with electron neutrino $|U_\mu|^2=|U_\tau|^2 = 0$, mixing with muon neutrino $|U_e|^2=|U_\tau|^2 = 0$ and mixing with tau neutrino $|U_e|^2=|U_\mu|^2 = 0$.
For the case of mixing with tau neutrino, we have to account for the sterile neutrinos with masses $M_N < M_\tau$ produced in tauon decays.
We have to consider additional processes \cite{Gorbunov:2007ak}:
\begin{eqnarray}
\label{eq:tau1}
\frac{d Br ( \tau^+ \to H^+ N)}{d E_N} & = & \tau_\tau |U_\tau|^2 \frac{G_F^2 |V_H|^2 f_H^2}{16 \pi} M^3_\tau \left( \left( 1 - \frac{M_N^2}{M_\tau^2} \right)^2 - \frac{M_H^2}{M_\tau^2} \left( 1 + \frac{M_N^2}{M_\tau^2}  \right) \right) \times \nonumber\\
&&\times\sqrt{\left( 1 - \frac{(M_H-M_N)^2}{M_\tau^2} \right) \left( 1 - \frac{(M_H+M_N)^2}{M_\tau^2} \right)} \times\nonumber\\
&&\times \delta \left( E_N - \frac{M_\tau^2 - M_H^2 + M_N^2}{2 M_\tau} \right),\\
\frac{d Br ( \tau^+ \to \rho^+ N)}{d E_N} & = & \tau_\tau |U_\tau|^2 \frac{G_F^2 |V_{ud}|^2 g_\rho^2}{8 \pi M_\rho^2} M^3_\tau \left( \left( 1 - \frac{M_N^2}{M_\tau^2} \right)^2 + \frac{M_\rho^2}{M_\tau^2} \left( 1 + \frac{M_N^2 - 2 M_\rho^2}{M_\tau^2}  \right) \right) \times \nonumber\\
&&\times\sqrt{\left( 1 - \frac{(M_\rho-M_N)^2}{M_\tau^2} \right) \left( 1 - \frac{(M_\rho+M_N)^2}{M_\tau^2} \right)} \times\nonumber\\
&&\times \delta \left( E_N - \frac{M_\tau^2 - M_\rho^2 + M_N^2}{2 M_\tau} \right),\\
\label{eq:tau3}
\frac{d Br ( \tau^+ \to \nu_\alpha l_\alpha^+ N)}{d E_N} & = & \tau_\tau |U_\tau|^2 \frac{G_F^2}{4 \pi^3} M_\tau^2 \left( 1 - \frac{M_l^2}{M_\tau^2 + M_N^2 - 2 E_N M_\tau}  \right)^2 \sqrt{E_N^2 - M_N^2} \times\nonumber\\
&&\times \Big( (M_\tau - E_N) \left(1 - \frac{M_N^2 + M_l^2}{M_\tau^2} \right) - \left( 1 - \frac{M_l^2}{M_\tau^2 + M_N^2 - 2 E_N M_\tau}  \right)\nonumber\\
&&\times \left( \frac{(M_\tau - E_N)^2}{M_\tau} +\frac{E_N^2-M_N^2}{3 M_\tau} \right) \Big),
\end{eqnarray}
here $H \in \{\pi^+, K^+ \}, \alpha \in \{e, \mu \}$.
Note that if sterile neutrino mixes not only with tau neutrino but with electron neutrino or muon neutrino as well, then another process $\tau \to \nu_\tau l_\alpha N$ becomes available.

To account for the processes \eqref{eq:tau1} -- \eqref{eq:tau3} one have to study the processes in which tauons are produced.
Tauons are produced in the decays of heavy mesons, the most significant ones being: $D_s^+ \to \tau^+ \nu_\tau$, $B^0 \to D^- \tau^+ \nu_\tau$, $B^0 \to D^{*-} \tau^+ \nu_\tau$, $B^+ \to \bar{D}^0 \tau^+ \nu_\tau$, $B^+ \to \bar{D}^{*0} \tau^+ \nu_\tau$ \cite{Tanabashi:2018oca}.
The branching ratios for the tauon production coincide with ones described by \eqref{Br2}, \eqref{Br3}, \eqref{Br_vec}, \eqref{eq:tau1} -- \eqref{eq:tau3} where sterile neutrino $N$ and lepton $l$ are replaced by tauon $\tau$ and active neutrino $\nu_\tau$.
One can obtain the resulting tauon momentum and coordinates with formulae \eqref{p_Ny} -- \eqref{eq:zN} in the same way.

To obtain $Br (H \to N...)$ one needs to integrate \eqref{Br2}, \eqref{Br3}, \eqref{Br_vec}, \eqref{eq:tau1} -- \eqref{eq:tau3} over $E_N$ from $M_N$ to $\frac{1}{2 M_H} \left(M_H^2 + M_N^2 -(M_{H'}+M_l)^2\right)$.
Note that other processes that may contribute to neutrino production have been studied \cite{Bondarenko:2018ptm}.
The most important of them is multi-meson decay channels.
For heavier sterile neutrino their importance becomes more significant.
By ignoring these states we can underestimate the total inclusive width $\frac{1}{\tau_H} \times \sum_X Br(H \to NX)$ by about 20$\%$ \cite{Bondarenko:2018ptm}.
On the other hand, as seen from Figs. \ref{fig:sensitivity-e}, \ref{fig:sensitivity-mu}, at $M_N \gtrsim2$ GeV too few sterile neutrinos reach Near Detector to be successfully registered.
For $M_N < 2$ GeV the multi-meson channels don't contribute significantly.

We note that many of the mentioned processes also prove to be insignificant in the interesting mass range.
To determine this we calculate the contribution of each process to sterile neutrino production.
If the process' branching value is less than $1\%$ for all considered values of sterile neutrino mass, then it is taken out of consideration.
This way the following processes were considered insignificant: $K^0_S \to \pi^+ l^- N$, $B^- \to \pi^0 l^- N$, $B^0 \to \pi^+ l^- N$, $B_s^0 \to K^+ l^- N$, $B^0 \to \rho^+ l^- N$, $B^- \to \rho^0 l^- N$, $B_s^0 \to K^{*+} l^- N$.
We note that for the case of mixing with tau neutrino only $D_s$ and heavier mesons can kinematically simultaneously produce the sterile neutrino and tauon in their decays (if $M_H>M_\tau+M_N $).

\end{document}